\documentclass[prb,aps,twocolumn,amstex,preprintnumbers,amsmath,amssymb,array,tabularx]{revtex4}
\usepackage{graphicx}
\usepackage{dcolumn}
\usepackage{bm}
\usepackage{lscape} 
\usepackage{amsmath}
\usepackage{amsxtra}
\usepackage{tabularx}

\DeclareGraphicsExtensions{.jpg,.pdf,.eps}

\begin{document}
\title{Edge Effects in Finite Elongated Graphene Nanoribbons}

\author{$\mbox{Oded Hod}^1$, $\mbox{Juan E. Peralta}^2$, and
  $\mbox{Gustavo E. Scuseria}^1$}

\affiliation{$^1$Department of Chemistry, Rice University, Houston,
  Texas 77005-1892;\\ $^2$Department of Physics, Central Michigan
  University, Mt. Pleasant, MI 48859}

\date{\today}

\begin{abstract}
  We analyze the relevance of finite-size effects to the electronic
  structure of long graphene nanoribbons using a divide and conquer
  density functional approach.  We find that for hydrogen terminated
  graphene nanoribbons most of the physical features appearing in the
  density of states of an infinite graphene nanoribbon are recovered
  at a length of $40$~nm.  Nevertheless, even for the longest systems
  considered ($72$~nm long) pronounced edge effects appear in the
  vicinity of the Fermi energy.  The weight of these edge states
  scales inversely with the length of the ribbon and they are expected
  to become negligible only at ribbons lengths of the order of
  micrometers.  Our results indicate that careful consideration of
  finite-size and edge effects should be applied when designing new
  nanoelectronic devices based on graphene nanoribbons.  These
  conclusions are expected to hold for other one-dimensional systems
  such as carbon nanotubes, conducting polymers, and DNA molecules.
\end{abstract}

\maketitle


Graphene nanoribbons (GNRs) have been suggested as potential
candidates for replacing electronic components and interconnects in
future nanoelectronic~\cite{Novoselov2004, Ezawa2006, Barone2006,
  Son2006-2, Han2007, Geim2007, Ozyilmaz2007} and
nanospintronic~\cite{Son2006, Hod2007, Hod2007-2} devices.
Experiments have revealed the possibility of obtaining a wide range of
electronic behavior when studying these systems, ranging from coherent
transport~\cite{deHeer2006} suitable for interconnects, to field
effect switching capabilities~\cite{Chen2007} needed for electronic
components design.  While many transport experiments involve long
segments of GNRs~\cite{Novoselov2004, Zhang2005-2, deHeer2006,
  Novoselov2007}, recent developments have allowed the fabrication of
graphene based quantum dots~\cite{Fujita2005, Han2007, Ozyilmaz2007,
  Vazquez2007}.  The reduced dimension of these quantum dots
introduces important physical phenomena such as quantum confinement
and {\em edge effects}.~\cite{Silvestrov2007, Shemella2007, Ezawa2007,
  Rossier2007, Jiang2007, Rudberg2007, Kan2007, Hod2007-2} Several
theoretical studies have emphasized the importance of such effects
when considering the electronic structure,~\cite{Rochefort1999,
  Wu2000, Lu2004, Li2005, Chen2006} electric
transport,~\cite{Anantram1998, Orlikowski2001, Compernolle2003,
  Jiang2005, Nemec2006} and magnetic~\cite{Okada2003, Chen2004-2,
  Chen2005} properties of finite carbon nanotubes (CNTs).  These
effects are expected to be manifested in experiments involving the
dielectric screening constants,~\cite{Lu2004} optical
excitations~\cite{Chen2004-1} and Raman spectrum~\cite{Saito1999} of
such systems.  Similar to CNTs it is predicted that the physical
characteristics of {\em finite} GNRs may be considerably different
from those of their {\em infinite} counterparts. Therefore, it is
essential to identify the limit at which finite-size effects have to
be taken into account.  An important question therefore arises: what
is the length at which a finite GNR becomes indistinguishable from its
infinite counterpart?
\begin{figure}[h]
  \begin{center}
    \includegraphics[width=8.8cm]{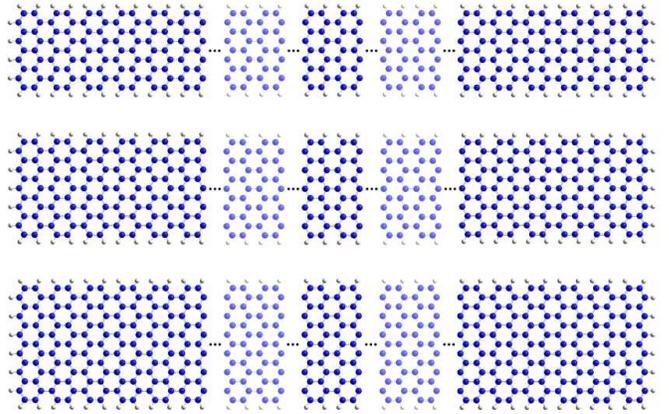}
  \end{center}
  \caption{Schematic representation of the 04- (upper panel), 05-
    (middle panel) and 06- (lower panel) finite armchair GNRs studied.
    Shown are the left and right terminating units, and the central
    part, which is replicated to produce the finite elongated system.}
  \label{Fig: GNR Geometries}
\end{figure}

The purpose of this Letter is to provide a quantitative answer to this
question based on first-principles calculations.  To this end, we
employ density functional theory (DFT) to study the electronic
structure of hydrogen terminated GNRs as a function of their length,
up to $72$~nm.  Finite-size effects are studied using a divide and
conquer approach for first-principles electronic structure and
transport calculations through finite elongated
systems.~\cite{Hod2006-2} A careful comparison with the electronic
structure of infinitely long periodic ribbons enables us to determine
the limit at which a finite GNR can be fairly approximated by its
infinite periodic counterpart.  Our results show that most of the
physical features appearing in the density of states (DOS) of the
infinite periodic system are recovered at a length of $40$~nm.
Nevertheless, pronounced features in the DOS resulting from edge
effects appear in the vicinity of the Fermi energy.  These features
are expected to persist up to ribbon lengths of the order of
micrometers.
\begin{figure}[h]
  \begin{center}
    \includegraphics[width=8.65cm]{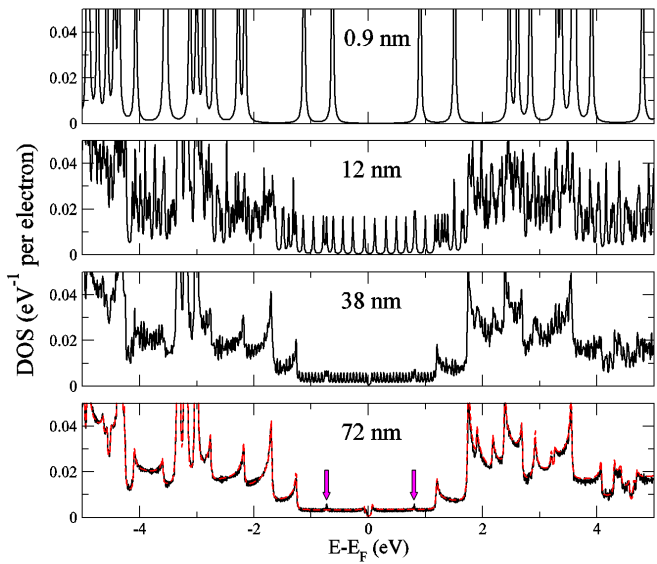}
  \end{center}
  \caption{Density of states of the hydrogenated 05-ACGNR calculated
    for several ribbon lengths and compared to the DOS of the periodic
    system (red dashed curve superimposed on the $72$~nm long 05-ACGNR
    DOS black curve in the lower panel).  The energy axis origin of
    all the panels is set to the Fermi energy of the periodic system
    ($-3.78$ eV). Edge states are indicated by purple arrows in the
    lower panel.}
  \label{Fig: DOS 05-ACGNR}
\end{figure}

A set of three finite armchair (AC) GNRs (ACGNRs) with consecutive
widths is considered.  We denote them as $N$-ACGNR, where $N$ stands
for the number of hydrogen atoms passivating the zigzag edges of the
terminating units (see Fig.~\ref{Fig: GNR Geometries}).  The relaxed
structures have been obtained using Pople's 3-21 Gaussian basis
set~\cite{Binkley1980} and the local spin density
approximation~\cite{Keywords} of density functional theory as
implemented in the {\it Gaussian} suite of programs.~\cite{Frisch2004}
The DOS calculations have been performed using the screened exchange
hybrid density functional of Heyd, Scuseria and Ernzerhof
(HSE06)~\cite{hse, hse-errata, Izmaylov2006} with the polarized
6-31G** Gaussian basis set.~\cite{Hariharan1973} The HSE06 functional
has been tested in a wide variety of materials and has provided good
agreement between predicted properties~\cite{Barone2006} of narrow
nanoribbons and measured values.~\cite{Han2007} The inclusion of
short-range exact-exchange in the HSE06 functional makes it suitable
to treat electronic localization effects~\cite{Kudin2002, Prodan2005}
which are known to be important in this type of
materials.~\cite{Kobayashi1993, Fujita1996, Nakada1996,
  Wakabayashi1999, Okada2001, Niimi2005, Lee2005, Son2006, Son2006-2,
  Kobayashi2006,Hod2007, Hod2007-2}

The DOS was calculated using the following relation:
\begin{equation}
  \rho(E)=-\frac{1}{\pi}\Im\{Tr[G^r(E)S]\},
  \label{Eq: DOS}
\end{equation}
where, $S$ is the overlap matrix, $G^r(E)=[\epsilon S-H]^{-1}$ is the
retarded Green's function (GF), $\epsilon=E+i\eta$, $E$ is the energy,
$H$ is the Hamiltonian matrix, and $\eta\rightarrow 0^+$ is a small
imaginary part introduced in order to shift the poles of the GF from
the real axis.  The Hamiltonian matrix is calculated using a divide
and conquer DFT approach.~\cite{Hod2006-2} Within this approach, $H$
is given in a localized basis set representation by a block
tridiagonal matrix,~\cite{Calzolari2004, Rocha2006, Adessi2006} where
the first and last diagonal blocks correspond to the two terminating
units of the ribbon (see Fig.~\ref{Fig: GNR Geometries}).  The
remaining diagonal blocks correspond to the central part of the GNR
which is composed of a replicated unit cell.  The terminating units
and the replicated central part unit cell are chosen to be long enough
such that the block-tridiagonal representation of $H$ (and $S$) is
valid.  The terminating units diagonal Hamiltonian blocks and their
coupling to the central part are evaluated via a molecular calculation
involving the two terminating units and one unit cell cut out of the
central part.  We approximate the replicated unit cell blocks of the
central part and the coupling between two such adjacent blocks to be
constant along the GNR, and extract them from a periodic boundary
conditions~\cite{Kudin2000} calculation.  The resulting
block-tridiagonal matrix $(\epsilon S-H)$ is then partially inverted,
using an efficient algorithm,~\cite{Godfrin1991} to obtain the
relevant GF blocks needed for the DOS calculation.~\cite{DaC} A
broadening factor of $\eta = 0.01$~eV was used.  In an experimental
setup this may correspond to broadening due to electron-phonon
coupling, surface/contacts effects, and finite temperature effects.

In Fig.~\ref{Fig: DOS 05-ACGNR} we present the DOS of the
quasi-metallic 05-ACGNR at an energy range of $\pm 5$~eV around the
Fermi energy of the infinite system for several ribbon lengths.  It
can be seen that for a $12$~nm ribbon (second panel from the top) the
DOS is composed of a set of irregularly spaced energy levels and is
uncorrelated with the DOS of the infinite GNR (dashed red curve in the
lower panel).  As the length of the GNR is increased the agreement
between the DOS of the finite and the periodic systems increases.  At
a length of $38$~nm one can clearly see the emergence of three
important characteristics of the DOS of the infinite 05-ACGNR, namely,
the buildup of the Van-Hove singularities, the constant DOS at the
vicinity of the Fermi energy, and the appearance of the DOS dip at the
Fermi energy.  When the length of the ribbon exceeds $70$~nm, apart
from the footprints of two minor edge states (indicated by purple
arrows in the lower panel) finite-size effects become negligible and
most of the physical features appearing in the DOS of the infinite
system are fully recovered.

\begin{figure}[h]
  \begin{center}
    \includegraphics[width=8.65cm]{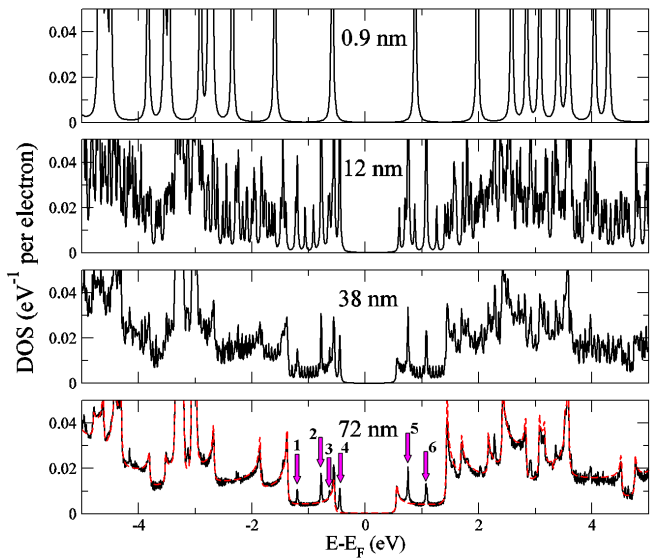}
  \end{center}
  \caption{Density of states of the hydrogenated 04-ACGNR calculated
    for several ribbon lengths and compared to the DOS of the periodic
    system (red dashed curve superimposed on the $72$~nm long 04-ACGNR
    DOS black curve in the lower panel).  The energy axis origin of
    all the panels is set to the Fermi energy of the periodic system
    ($-3.75$~eV). Edge states are indicated by purple arrows in the
    lower panel.}
  \label{Fig: DOS 04-ACGNR}
\end{figure}

We now turn to discuss the semi-conducting 04-, and 06-ACGNRs.
Similar to the quasi-metallic case, we consider the Van-Hove
singularities and the energy gap as two important characteristic
features of the DOS.  In Fig.~\ref{Fig: DOS 04-ACGNR} the DOS of the
finite 04-ACGNR is presented for several CNT lengths at a region of
$\pm 5$~eV around the Fermi energy of the infinite periodic system.
The energy gap is well captured even for the $0.9$~nm ribbon, while
the reconstruction of the Van-Hove singularities as the length of the
finite semi-conducting GNR is increased, follows the same lines
described above for the quasi-metallic case.  Nonetheless, even for
the longest system studied ($72$~nm), considerable edge states appear
in the vicinity of the Fermi energy.  An indication of the local edge
nature of these states is that their weights in the total DOS plot
scale inversely with the length of the ribbon.  For the $72$~nm
04-ACGNR, the height of the edge states peaks is comparable to that of
the Van-Hove singularities corresponding to the valence band top and
the conductance band bottom.  Using simple extrapolation arguments one
can conclude that these edge states are expected to be experimentally
detectable even for micrometer long armchair nano ribbons.  Similar
features are obtained for the semiconducting 06-ACGNR (not shown).

\begin{figure}[h]
  \begin{center}
    \includegraphics[width=8.65cm]{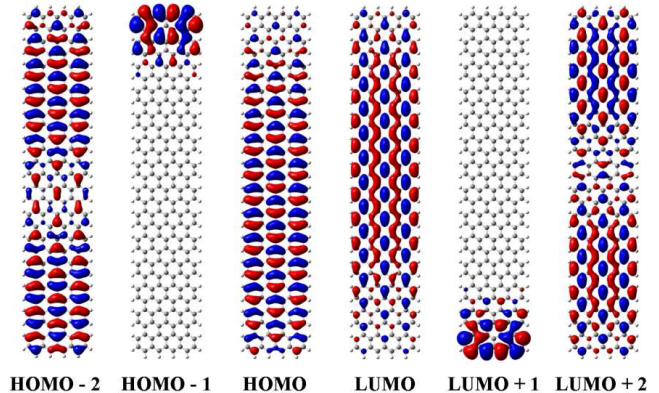}
  \end{center}
  \caption{Isosurface plots of several of the $\alpha$ spin component
    Kohn-Sham orbitals in the vicinity of the HOMO-LUMO gap of a
    $6$~nm long 04-ACGNR.  The orbital energies with respect to the
    Fermi energy of the infinite system ($-3.75$~eV) are $-0.83$,
    $-0.63$, $-0.60$, $0.68$, $0.75$, and $0.93$~eV for the HOMO-2,
    HOMO-1, HOMO, LUMO, LUMO+1, and LUMO+2 orbitals, respectively. The
    $\beta$ spin orbitals plots (not shown) are a mirror image of the
    $\alpha$ component plots with respect to a reflection plane
    perpendicular to the armchair edge and crossing the ribbon at its
    center of mass.  The isovalue is: $0.0075$}
  \label{Fig: Wave functions}
\end{figure}

To better understand the nature of these pronounced edge states, one
should note that unlike infinite ACGNRs the finite systems have zigzag
edges which may present magnetic ordering.~\cite{Kobayashi1993,
  Fujita1996, Nakada1996, Wakabayashi1999, Okada2001, Niimi2005,
  Lee2005, Son2006, Son2006-2, Kobayashi2006, Hod2007, Hod2007-2} In
Fig.~\ref{Fig: Wave functions} we show the single particle Kohn-Sham
orbitals appearing in the vicinity of the Fermi level of a $6$~nm long
04-ACGNR.  The highest occupied molecular orbital (HOMO) and the
lowest unoccupied molecular orbital (LUMO) present a delocalized state
with a major portion of the orbital spanned on the central part of the
ribbon.  Due to their extended nature, the energy of such states will
depend on the length of the ribbon.  Therefore, it may change
considerably when going from very short ribbons, such as the $0.9$~nm
system presented in the upper panel of Fig.~\ref{Fig: DOS 04-ACGNR} to
an infinite system where they contribute to the Van-Hove singularities
corresponding to the top of the valence band and the bottom of the
conduction band.  This is especially true for the LUMO state which
presents a longitudinal characteristic length larger than the full
extent of the ribbon.  The HOMO-1 and LUMO+1 orbitals are of clear
localized nature confined to a small region near the edge of the
ribbon.  The orbital energies of these states (see caption of
Fig.~\ref{Fig: Wave functions}) correspond well to the edge states
marked as $3$ and $5$ in the lower panel of Fig.~\ref{Fig: DOS
  04-ACGNR}.  As can be seen, due to their localized zigzag edge
nature, these states appear in all lengths scales studied ranging from
$0.9$~nm up to $72$~nm, and are expected to persist even at the
microscale.  Similar to the HOMO and LUMO orbitals, the HOMO-2 and
LUMO+2 states are of an extended character.  Nevertheless, from a
careful examination of these states it can be seen that they are
divided into two extended regions each corresponding to a different
zigzag edge of the ribbon.  Therefore, they can be viewed as extended
edge states.  As mentioned above, the energy of such states will
depend on the length of the ribbon, corresponding roughly to the DOS
peaks marked as $2$ and $6$ in the lower panel of Fig.~\ref{Fig: DOS
  04-ACGNR}, and not observed in the DOS of the $0.9$~nm system.  We
could not locate a Kohn-Sham orbital of the $6$~nm long system that
corresponds to edge state $4$ appearing in the lower panel of
Fig.~\ref{Fig: DOS 04-ACGNR}.  This suggests that similar to the
HOMO-2 and LUMO+2 states, this orbital is of extended edge nature and
appears at length scales longer than $6$~nm.

Edge effects play a dominant role when considering the electronic and
magnetic character of low dimensional elongated systems.  This has
been shown for systems such as graphene
nanoribbons,~\cite{Kobayashi1993, Fujita1996, Nakada1996,
  Wakabayashi1999, Okada2001, Niimi2005, Lee2005, Son2006, Son2006-2,
  Kobayashi2006, Hod2007, Hod2007-2} carbon
nanotubes,~\cite{Anantram1998, Wu2000, Orlikowski2001,
  Compernolle2003, Okada2003, Chen2004-2, Li2005, Jiang2005, Chen2005,
  Chen2006, Nemec2006} and other related structures.~\cite{Okada2001}
When considering the study of graphene nanoribbons as candidates for
future nano-electronic devices, it is important to identify the
contribution of finite-size effects to the physical properties of the
entire system.  In the present letter, we have shown that while most
of the physical features characterizing an infinitely long ACGNR are
recovered for ribbons $\sim 40$~nm long, prominent edge effects can be
present up to ribbon lengths as high as a few micrometers.  This has
been done by studying the DOS of GNRs as a function of their length
and comparing with that of the periodic system.  Even though the exact
details of such effects are expected to depend on the nature of the
terminating units, care must be taken regarding their influence on the
electronic character of the system.  Our conclusions largely hold for
other finite elongated systems such as carbon nanotubes, conducting
polymers, and DNA molecules, as well.  Research along these lines is
currently in progress.

This research was supported by the National Science Foundation under
Grant CHE-0457030 and the Welch foundation.  J.E.P. acknowledges the
support from the President's Research Investment Fund from Central
Michigan University.  O.H. would like to thank the generous financial
support of the Rothschild and Fulbright foundations.  Part of the
computational time employed in this work was provided by the Rice
Terascale Cluster funded by NSF under Grant EIA-0216467, Intel, and
HP, and by the ADA cluster that is supported by a Major Research
Infrastructure grant from the National Science Foundation
(CNS-0421109), Rice University and partnerships with AMD and Cray.

\bibliographystyle{./prsty.bst} \bibliography{GNR-DOS}

\newpage
\end{document}